\documentclass[conference]{IEEEtran}
\IEEEoverridecommandlockouts
\usepackage{cite}
\usepackage{amsmath,amssymb,amsfonts}
\usepackage{algorithmic}
\usepackage{algorithm}
\usepackage{graphicx}
\usepackage{textcomp}
\usepackage{xcolor}
\usepackage{cite}
\usepackage{bm}
\usepackage{bbm}
\usepackage{graphics}
\usepackage{color}
\usepackage{subcaption}
\usepackage{dsfont}
\usepackage{caption}

\usepackage{MnSymbol}

\newcommand{\vecelem}[2]{\errmessage{DO NOT USE}}    
\usepackage{color}

\include{def}
\def\BibTeX{{\rm B\kern-.05em{\sc i\kern-.025em b}\kern-.08em
    T\kern-.1667em\lower.7ex\hbox{E}\kern-.125emX}}

\def\tW{\pmb{\mathcal{W}}}
\def\a{\bm{\mathbf{a}}}

\def\c{\bm{\mathbf{c}}}
\def\A{\bm{\mathbf{A}}}
\def\B{\bm{\mathbf{B}}}
\def\C{\bm{\mathbf{C}}}
\def\I{\bm{\mathbf{I}}}
\def\D{\bm{\mathbf{D}}}

\def\J{\bm{\mathbf{J}}}

\def\Q{\bm{\mathbf{Q}}}
\def\R{\bm{\mathbf{R}}}
\def\S{\bm{\mathbf{S}}}
\def\U{\bm{\mathbf{U}}}
\def\V{\bm{\mathbf{V}}}

\def\u{\bm{\mathbf{u}}}
\def\c{\bm{\mathbf{c}}}

\def\f{\bm{\mathbf{f}}}
\def\e{\bm{\mathbf{e}}}
\def\x{\bm{\mathbf{x}}}

\def\s{\bm{\mathbf{s}}}
\def\0{\bm{\mathbf{0}}}

\def\Toep{\mathrm{Toeplitz}}

\begin{document}

\title{Accurate DOA Estimation Based on Real-Valued Singular Value Decomposition\\
\thanks{This work was supported by Fundamental Research Funds for Central Universities (Grand No. WUT: 2018 IVA 097).}
}

\author{\IEEEauthorblockN{ Hui Cao}
\IEEEauthorblockA{\textit{School of Information Engineering} \\
\textit{Wuhan University of Technology}\\
Wuhan, China \\
iehuihcao@whut.edu.cn}
\and
\IEEEauthorblockN{Qi Liu}
\IEEEauthorblockA{\textit{Department of Electronic Engineering} \\
\textit{City University of Hong Kong}\\
Hong Kong, China \\
lq$\_$skyven@126.com}
}

\maketitle

\begin{abstract}
In this paper, an accurate direction-of-arrival (DOA) estimator is developed based on the real-valued singular value decomposition (SVD) of covariance matrix. Unitary transform on the complex-valued covariance matrix is first applied, and then SVD performs on the resulting real-valued data matrix. The singular vector is then utilized with a weighted least squares (WLS) method to achieve DOA estimation. The performance of the proposed algorithm is compared with several state-of-the-art methods as well as the CRB. The results indicate the accuracy and effectiveness of the proposed method.
\end{abstract}


\section{Introduction}
In array signal processing, direction-of-arrival (DOA) estimation of multiple narrowband sources plays an important role and it has wide applications in radar, sonar, radio astronomy, and mobile communications \cite{Krim1996,Stoica1990, JLi2018, Liu2017, Lin2017, Khab2014}. Many high resolution methods have been developed over the past decades, such as the root-MUSIC \cite{Rao1989}, and MODE \cite{Stoica1990, Stoica19901}, which attain different trade-offs between accuracy and complexity of the estimation. With the use of a unitary transform, many real-valued method have been proposed to reduce the computational complexity, such as the unitary ESPRIT \cite{Haardt1995} method, which exploits the centro-Hermitian property of the forward-backward covariance matrix to obtain real-valued data matrix. \cite{Pesa2000} considers a real-valued (unitary) formulation of the root-MUSIC DOA (U-root-MUSIC) estimation technique while \cite{Yan2018} presents a second forward-backward (SFB) averaging technique to transform the real symmetrical covariance matrix of the U-root-MUSIC into a real bisymmetrical one.

In this paper, we propose a subspace method based on a real-valued singular value decomposition (SVD), the singular vectors obtained from the SVD is utilized in a weighted least squares (WLS) iteration process to achieve the estimation of the DOA.

The rest of the paper is organized as follows. In Section~\ref{sec:alg dev}, the DOA estimation problem is first formulated. Then using a unitary transform, a real-valued SVD is performed to receive the subspace of the received signal.  DOA estimation is achieved with the use of the singular vector and a WLS iteration procedure. In Section~\ref{sec:numr res}, numerical examples are included to evaluate the estimation performance of the proposed algorithm by comparing with several existing methods as well as the CRB. Finally, conclusions are drawn in Section~\ref{sec:concl}.

\section{Algorithm Development}\label{sec:alg dev}
Consider a ULA be composed of $M$ sensors, and it receives $K (K<M)$ narrowband sources imping from the directions $\theta_1,\cdots,\theta_K$. We assume that $N$ snapshots $\mathbf{x}(1),\mathbf{x}(2),\cdots,\mathbf{x}(N)$ are available. Then the $M\times 1$ vector is modeled as \cite{Rao1989}
\begin{align}\label{equ11D}
\x(t)=\A \s(t)+\mathbf n(t),
\end{align}
where $\A=[\a(\theta_1 ),\a(\theta_2 ),\cdots,\a(\theta_K )]$ is the $M\times K$ matrix of the signal direction vectors, $\s(t)$ is the vector of waveforms, $\mathbf n(t)$ is white sensor noise with variance $\sigma^2$, and
\begin{equation}\label{equ21D}
\mathbf{a}(\theta_k)=\left[1,e^{j(\frac{2\pi}{\lambda}d\sin\theta_k)},\cdots,e^{j(\frac{2\pi}{\lambda}d(M-1)\sin\theta_k)} \right]^T, k=1,\cdots,K
\end{equation}
is the $M\times 1$ steering vector. $^T$ denotes the transpose operator. Here, $d$ is the array interelement spacing and $\lambda$ is the wavelength, and let $d=\lambda/2$.

The conventional estimate of the covariance matrix is defined as
\begin{align}\label{equ31D}
\R &=\mathbb{E}[\x(t)\x^{H}(t)]=\A \S\A ^{H}+\sigma^2 {\mathbf I_{M}},
\end{align}
where $\mathbb{E}$ is the expectation operator, $^H$ is the Hermitian transpose operator, $\S_{K\times K}=\mathbb{E}[ \s(t) \s^{H}(t)]$ is source waveform covariance matrix. $I_{M}$ is identity matrix of size $M\times M$.
As is known, the covariance matrix $\R$ is a complex-valued matrix, to reduce the computation complexity, we can convert it into a real-valued covariance $\C$ using the unitary transform \cite{Haardt1995}
\begin{equation}\label{equrealC}
\C=\frac{1}{2}\Q_{M}^{H}(\R+\J_M\R^*\J_M)\Q_M=\Re\{\Q_{M}^{H}\R\Q_M\},
\end{equation}
where $^*$ denotes the conjugate operator, $\Re$ represents the real part, $\J_M$ is an $M\times M$ exchange matrix with ones on its antidiagonal and zeros otherwise. The unitary transformation matrix $\Q_M$ is defined as \cite{Haardt1995}:
\begin{equation}\label{equQ}
\Q_{M}=\left\{\begin{matrix}
\frac{\sqrt{2}}{2}\begin{bmatrix}
 \I_{l} & j \I_{l} \\
 \J_{l} & j/ \J_{l}
\end{bmatrix},\mathrm{for}~ M=2l\\
 \frac{\sqrt{2}}{2}\begin{bmatrix}
 \mathbf{I}_{l}& \mathbf{0}_{l} & j \mathbf{I}_{l}\\
\mathbf{0}_{l}^{T} & \sqrt(2) & \mathbf{0}_{l}^{T}\\
 \mathbf{J}_{l}& \mathbf{0}_{l} &-j/ \mathbf{J}_{l}
\end{bmatrix},\mathrm{for}~ M=2l+1,
\end{matrix}\right.
\end{equation}
where $\0_l$ is an $l\times 1$ zero vector.
Perform the SVD on the real-valued covariance matrix $\C$,
\begin{equation}\label{equ71D}
\C=\U_s\Lambda_s\V_s^H+\U_n\Lambda_n\V_n^H,
\end{equation}
where $\U_s$ and $\U_n$ are the signal and noise subspace, respectively. Both of them are real-valued matrix. To obtain the estimate of DOA, we can directly apply subspace method like root-MUSIC \cite{Rao1989,Pesa2000} or Unitary ESPRIT \cite{Haardt1995} with the use of signal/noise subspace. On the other hand, we can use the unitary transformation matrix $\Q_M$ to convert the subspace matrix $\U_s$ into a complex-valued one \cite{Pesa2000}, denoted as $\U_c$
\begin{equation}\label{equ81D}
\U_c=\Q\U_s,
\end{equation}
where $\U_c$ is a complex-valued singular vector contains the information of DOA. Denoting $\U_c = \begin{bmatrix} \u_1 & \u_2 & \cdots & \u_{M} \end{bmatrix}^{\mathrm{T}}$. Since there are $K$ sources, ${\mathrm{ rank}}(\U_c) = K$,  both $\u_i$, $i \in \{K+1, \cdots, M\}$ are $\0$ in noise-free case, where $\0\in \mathbb{C}^{M \times 1}$ denotes a vector with all 0 values.

The elements along the columns of $\U_{c}$ satisfy the LP property \cite{Chan2012}, that is:
\begin{eqnarray}\label{eq6}
&&\sum_{i=0}^K c_k [u_{k}]_{m-i} = 0, \\\nonumber
&& c_0=1,~k=1,\cdots,K,~m=K+1,\cdots,\cdots,M,
\end{eqnarray}
where $c_i=e^{j(\frac{2\pi}{\lambda}d\sin\theta_i)}$ is the LP coefficients, from which the LP coefficients are given by $K$ roots of the following polynomial \cite{Chan2012}:
\begin{align}
1+ \sum_{k=1}^K c_k z^{K-k} = 0,  \label{equ71}
\end{align}
where $z=e^{j(\frac{2\pi}{\lambda}d\sin\theta_k)}, k=1,\cdots,K$.
Define a Toeplitz matrix $\B$ as:
\begin{eqnarray}\small
\B &=& \Toep\left(\begin{bmatrix} c_K & \0_{1 \times
(M-K-1)} \end{bmatrix}^T,\right.\\\nonumber
&&\left. \begin{bmatrix} c_K & c_{K-1} & \cdots & c_1 & 1 & \0_{1 \times
(M-K-1)} \end{bmatrix}\right),
\end{eqnarray}
then \eqref{equ71} is rewritten as:
\begin{align}
\B \u_k = \D_k \c - \f_k = \0 , \quad k=1,2,\cdots,K,  \label{eq9}
\end{align}
where
\begin{align}
&\D_k = \begin{bmatrix}\notag
 [\u_{k}]_{K} & [\u_{k}]_{K-1} & \cdots & [\u_{k}]_1 \\
 [\u_{k}]_{K+1} & [\u_{k}]_K & \cdots & [\u_{k}]_2 \\
 \vdots & \vdots & \ddots & \vdots \\
 [\u_{k}]_{M-1} & [\u_{k}]_{M-2} & \cdots & [\u_{k}]_{M-K}
 \end{bmatrix},
 \end{align}
 \begin{align}\notag
&\c = \begin{bmatrix} c_1 & c_2 & \cdots & c_K \end{bmatrix}^T, 
 \end{align}
 and
  \begin{align}\notag
&\f_k = - \begin{bmatrix} [\u_{k}]_{K+1} & [\u_{k}]_{K+2} & \cdots & [\u_{k}]_M \end{bmatrix}^T.
\end{align}

Collecting all $K$ vectors in \eqref{eq9} together, it yields:
\begin{equation}\label{equ12}
[(\B\u_1)^T \quad (\B\u_2)^T \quad \cdots \quad (\B\u_K)^T ]^T = \D \c - \f = \0,
\end{equation}
with $\D = [ \D_1^T \quad\D_2^T \quad \cdots \quad \D_K^T ]^T$, $\f = [ \f_1^T \quad \f_2^T \quad \cdots \quad \f_K^T ]^T$. It is easy to calculate $\c$ from \eqref{equ12} without the existence of noise.

In the presence of noise, $(\D \c - \f)$ in \eqref{equ12} does not equal a zero vector. We define the result of $(\D \c - \f)$ as $\e \in \mathbb{C}^{(M-K)K\times 1}$. The resulting problem is converted to find $\c$ from $\D \c - \f = \e$, which can be solved by the following WLS minimization \cite{Chan2012} :
\begin{align}
&\hat{\c} = \arg \min_{\c} \e^H \tW \e
=\left(\D^H \tW \D\right)^{-1}\D^H \tW \f,  \label{eq17}
\end{align}
where $\tW$ is a symmetric weight matrix and its optimal choice is derived with the covariance matrix w.r.t. $\e$, given by
\begin{align}
\tW = \sigma^2 \left[ \mathds{E} \left\{ \e \e^H \right\}\right]^{-1} = \I_K \otimes (\A \A^H)^{-1}.  \label{eq19}
\end{align}

Next, substituting $\hat{\c}$ in \eqref{equ71}, and solving for the roots, denoted
by $\hat{u}_k$. Then the DOA estimate $\hat{\theta}_k$ is:
\begin{align}\label{equ91D}
\hat{\theta}_k &=\sin^{-1}(\frac{\lambda}{2\pi d}\hat{\mu}_k).
\end{align}
We summarise our proposed algorithm in Table~\ref{tabl1}.
\begin{table}[htbp]
\caption{Proposed algorithm }
\label{tabl1}
\begin{tabular}{l}
  \hline
  \\
Step 1: Compute the covariance matrix $\R$ using received signal $\x$.\\
Step 2: Construct real-valued covariance matrix $\C$ using unitary matrix $\Q_M$.\\
Step 3: Perform SVD on real-valued covariance matrix $\C$.\\
Step 4: Use the singular vector to construct $\U_c$ and utilize the WLS \\
~~~~~~~~~ iteration procedure in \eqref{eq17}, then obtain the estimate of DOA with \eqref{equ91D}.\\
\\
  \hline
\end{tabular}
\end{table}

\section{Numerical Results}\label{sec:numr res}
In this section, simulations are conducted to evaluate the performance of the proposed algorithm on DOA estimation, We compare the proposed algorithm with root-MUSIC \cite{Rao1989}, unitary ESPRIT \cite{Haardt1995}, MODE \cite{Stoica1990, Stoica19901} and SFB-U-root-MUSIC \cite{Yan2018} in the simulation. The CRB \cite{Stoica1990} is plotted as a benchmark. Assume the power of signals are the same, and is $\sigma_s^2$. The signal-to-noise ratio (SNR) is defined as $\mathrm{SNR}=10\log (\sigma_s^2/\sigma^2)$. We scale the noise power to produce different SNR conditions. The root mean square error (RMSE)
\begin{align}
\rm {RMSE} =\sqrt{ \mathbb{E} \{(\hat{\theta}_k-\theta_k)^2\} } \notag
\end{align}
is applied to measure the effectiveness of the proposed method. Each simulation results are tested with $200$ Monte Carlo trials, on the MATLAB R2017b of laptop with 32 GB RAM and 64-bit Windows 10 operating system.

\begin{figure}[htp!]
\centering
 \captionsetup{font={scriptsize}}
       \includegraphics[width=8cm]{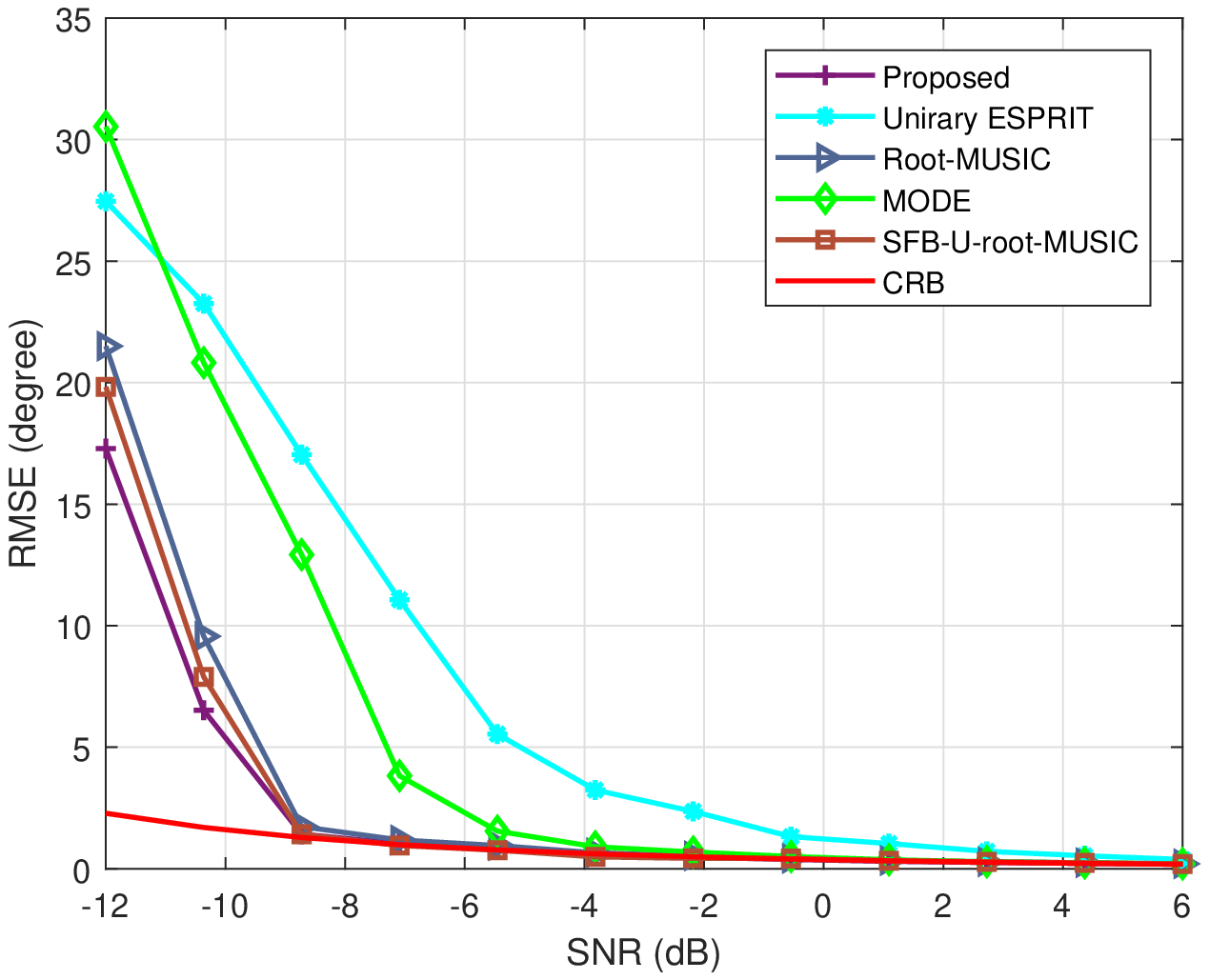}
        \caption{ RMSE of DOA estimates versus SNR. }
        \label{fig1}
    \end{figure}

In the simulation, two independent narrowband Gaussian signals are assumed to imping on a ULA from directions $[\theta_1~\theta_2]=[6^{\circ}, 45^{\circ}]$, where sensors number $M=10$, snapshots number $N=50$. We first investigate the RMSE of the proposed method versus SNR. As shown in Fig. \ref{fig1}, our proposed method has better threshold behavior, it attains the CRB at SNR=-8 dB and has lower RMSE than others at low SNR conditions.

\begin{figure}[htp!]
\centering
 \captionsetup{font={scriptsize}}
       \includegraphics[width=8cm]{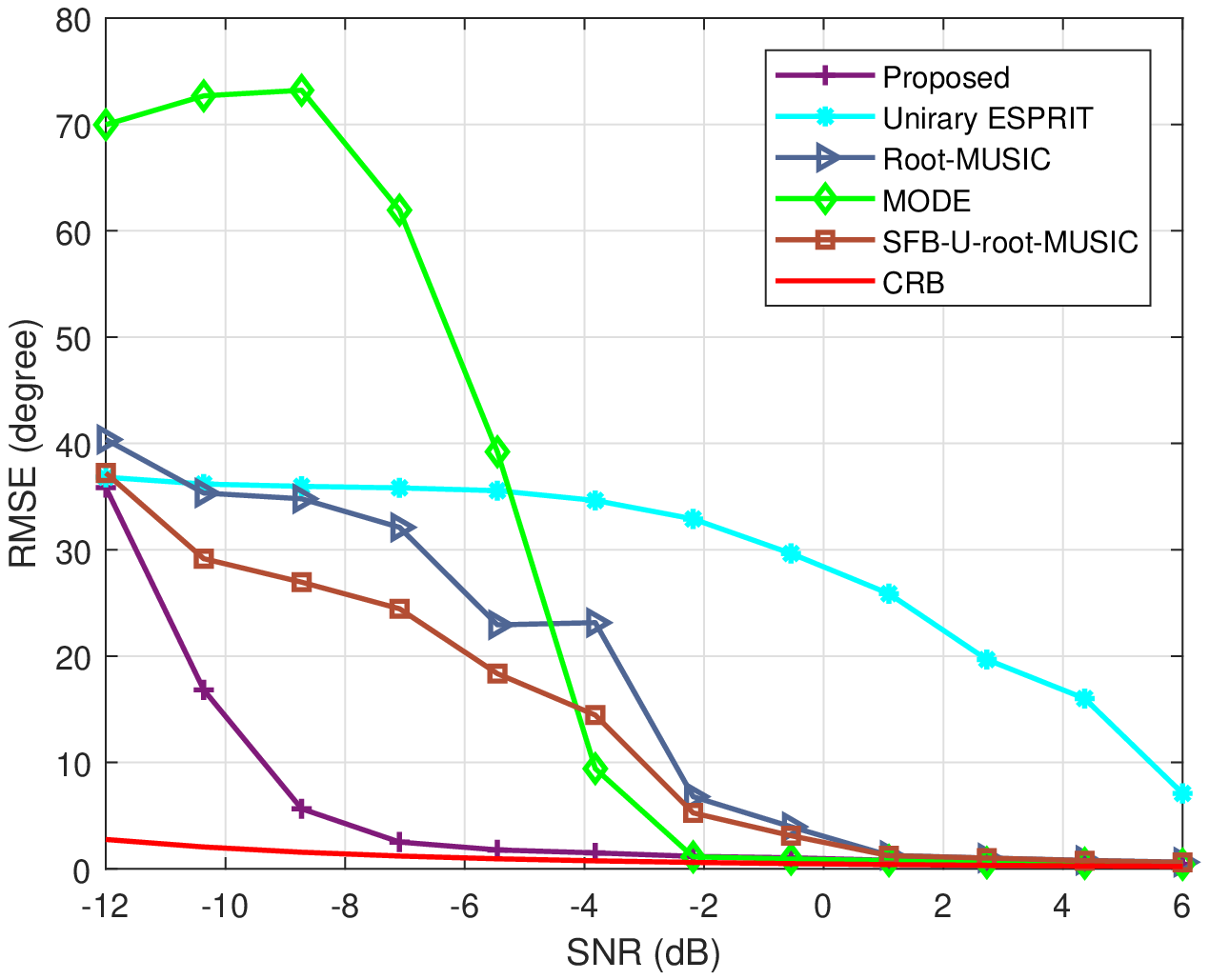}
        \caption{ RMSE of DOA estimates versus SNR. }
        \label{fig0}
    \end{figure}

To further investigate the performance of the proposed method we change the two DOA angles to $[\theta_1~\theta_2]=[30^{\circ}, 45^{\circ}]$. As shown in Fig. \ref{fig0}, in this case, our proposed method has better threshold behavior than all the other methods, while the Unitary ESPRIT method fails to work.

\begin{figure}[htp!]
\centering
 \captionsetup{font={scriptsize}}
       \includegraphics[width=8cm]{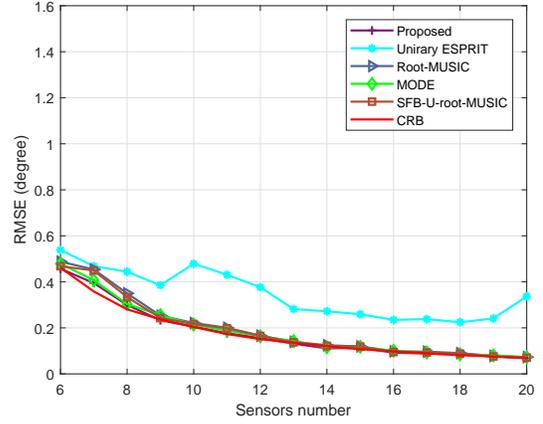}
        \caption{RMSE of DOA estimates versus array number. }
        \label{fig2}
    \end{figure}

\begin{figure}[htp!]
\centering
 \captionsetup{font={scriptsize}}
       \includegraphics[width=8cm]{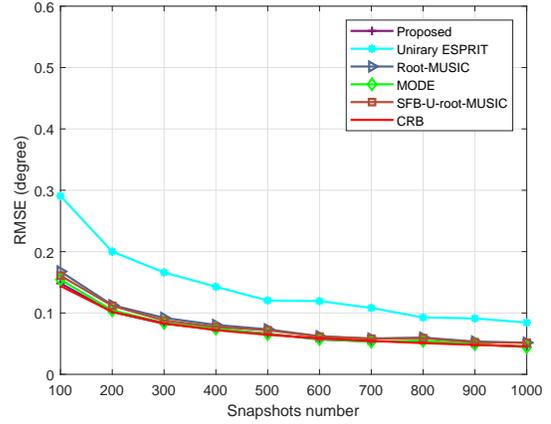}
        \caption{RMSE of DOA estimates versus snapshots number. }
        \label{fig3}
    \end{figure}

In Figs. \ref{fig2} and \ref{fig3}, we plot the RMSE of the proposed method and other methods by changing sensors number $M$ and snapshots number $N$. We fixed the SNR at 5 dB. We use the same data settings as in simulation one except the sensors number $M$ and snapshots number $N$. The sensors number $M$ varies from 6 to 20 in Fig. \ref{fig2}, while the snapshots number $N$ varies from 100 to 1000 in Fig. \ref{fig3}. It is seen that, in these two figures, our proposed method always attain the CRB and outperforms the unitary ESPRIT method, which again indicates the accuracy of the propose method.

\begin{figure}[htp!]
\centering
 \captionsetup{font={scriptsize}}
       \includegraphics[width=8cm]{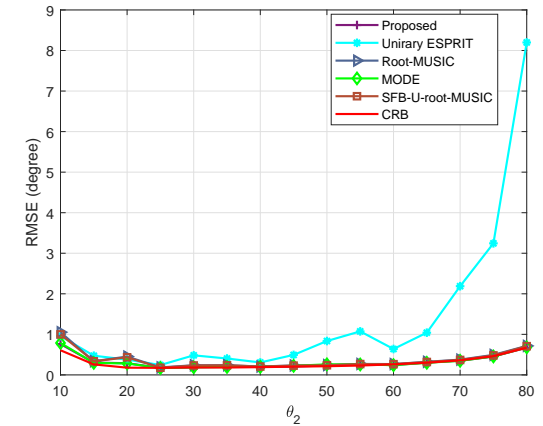}
        \caption{RMSE of DOA estimates versus changing $\theta_2$. }
        \label{fig4}
    \end{figure}

In the last experiment, we investigate the resolution ability of the proposed method by fixing one DOA angle and changing the other one. We use the same data settings as in simulation 1 but changes $\theta_2$ from $10^{\circ}$ to $80^{\circ}$. As seen in Fig. \ref{fig4}, our proposed method has similar performance with other methods and outperforms the Unitary ESPRIT method.


\clearpage

\section{Conclusion}\label{sec:concl}
An accurate DOA estimator is designed in this paper. The covariance matrix of signal is first transformed into a real-valued matrix with the use of a unitary transform matrix. Then the SVD of the real-valued covariance matrix is performed to reduce the complexity compare with direct decomposition on the complex-valued covariance matrix. The obtained singular vector is utilized to estimate the DOA with the use of a WLS iteration procedure. Simulation results verifies the effectiveness of the proposed method and shows that it provides more accurate DOA estimation at lower SNRs.



%
%

\end{document}